\documentclass[final,5p,times,twocolumn]{elsarticle}
\usepackage{amsmath,amssymb,amsfonts}
\usepackage{algorithmic}
\usepackage{graphicx}
\usepackage{epstopdf}
\usepackage{multirow}
\usepackage{booktabs}
\usepackage{footnote}
\usepackage{subfigure}
\usepackage{makecell}
\usepackage{color,bm}
\usepackage{algorithm}
\usepackage{geometry}
\definecolor{v1}{RGB}{0,0,0}
\definecolor{v2}{RGB}{0,0,0}
\definecolor{v3}{RGB}{0,0,0}
\definecolor{v4}{RGB}{0,0,0}

\journal{Control Engineering Practice}

\begin{document}

\begin{frontmatter}



\title{Bayesian Optimization Assisted Meal Bolus Decision Based on Gaussian Processes Learning and Risk-Sensitive Control\tnotemark[1]}

\tnotetext[1]{This work was supported in part by the National Natural Science Foundation of China under Grant 61973030, and in part by the Peking University People's Hospital Scientific Research Development Funds RDY2019-05.}



\author[label1]{Deheng Cai}
\author[label2]{Wei Liu}
\author[label2]{Linong Ji}
 \author{Dawei Shi\corref{cor1}\fnref{label1}}
 \cortext[cor1]{Corresponding author}
 \ead{daweishi@bit.edu.cn}

 \address[label1]{State Key Laboratory of Intelligent Control and Decision of Complex Systems, School of Automation, Beijing Institute of Technology, Beijing, China}
     \address[label2]{Department of Endocrine and Metabolism, Peking University People's Hospital, Beijing, China}
\begin{abstract}
Effective postprandial glucose control is important to glucose management for subjects with diabetes mellitus. In this work, a data-driven meal bolus decision method is proposed without the need of subject-specific glucose management parameters. The postprandial glucose dynamics is learnt using Gaussian process regression. Considering the asymmetric risks of hyper- and hypoglycemia and the uncertainties in the predicted glucose trajectories, an asymmetric risk-sensitive cost function is designed. Bayesian optimization is utilized to solve the optimization problem, since the gradient of the cost function is unavailable. The proposed approach is evaluated using the 10-adult cohort of the FDA-accepted UVA/Padova T1DM simulator and compared with the standard insulin bolus calculator. For the case of announced meals, the proposed method achieves satisfactory and similar performance in terms of mean glucose and percentage time in [70, 180] mg/dL without increasing the risk of hypoglycemia. Similar results are observed for the case without the meal information (assuming that the patient follows a consistent diet) and the case of basal rate mismatches. In addition, advisory-mode analysis is performed based on clinical data, which indicates that the method can determine safe and reasonable meal boluses in real clinical settings. The results verify the effectiveness and robustness of the proposed method and indicate the feasibility of achieving improved postprandial glucose regulation through a  data-driven optimal control method.
\end{abstract}




\end{frontmatter}







\section{Introduction}
Diabetes mellitus (DM) is a chronic metabolic disease that is characterized by absolute insulin deficiency (type 1 diabetes (T1D)) or relative lack of insulin secretion and sensitivity (type 2 diabetes (T2D)). Patients with DM tend to suffer from the long term complications, e.g., retinopathy and nephropathy, due to poor blood glucose (BG) managements \cite{diabetes1993effect}. Nowadays, external insulin administration through a basal-bolus strategy with multiple daily injections (MDI) or continuous subcutaneous insulin infusion (CSII) pump is a popular way among the patients to control BG  \cite{Schiffrin1982}. With improved modern diabetes technologies, a closed-loop control system for BG regulations, named artificial pancreas (AP), is further developed through integrating the pumps and continuous glucose monitoring (CGM) sensors. {\color{v4}An AP automatically delivers insulin to achieve desired BG levels, based on CGM-driven feedback control algorithms \cite{Doyle2014,8263484,BOIROUX2017,INCREMONA2018}}. 

However, for the therapies above, the effective control of postprandial glucose still remains a challenge \cite{El2018}. At present, the feedforward control action in terms of preprandial insulin bolus is widely adopted to counteract the hyperglycemia associated with meal intakes. Specifically, a meal bolus is determined by a bolus calculator according to carbohydrate content of the meal, the BG level and subject-specific settings (carbohydrate ratio (CR) and correction factor (CF) profiles). Multiple efforts have been devoted to improve the performance of the bolus advisors. Considering the repetitive nature of the daily activities of the patient, Owens \emph{et al.} \cite{Run2006} proposed a Run-to-Run (R2R) algorithm to update the insulin bolus amount and timing daily. 
In Schiavon \emph{et al.} \cite{Schiavon2018}, a novel optimization method for CR was developed based on a validated index of insulin sensitivity estimated form CGM and CSII data. Combining case-based reasoning (CBR) with R2R, an advanced insulin bolus advisor through adapting CR and CF was presented in Herrero \emph{et al.} \cite{Herrero20151}. Similarly, Torrent-Fontbona \cite{Torrent-Fontbona2019} investigated a bolus insulin recommend system based on CBR, but provided a new reuse, revise and retain mechanism to adapt CR and CF. Besides CBR, other artificial intelligence techniques including fuzzy logic \cite{Liu2013} and reinforcement learning (RL) \cite{Sun2019} have been investigated for the bolus {\color{v3}advisors}. Moreover, several algorithms have also been explored to calculate the required bolus within the framework of an AP. {\color{v4}In Turksoy \emph{et al.} \cite{TURKSOY2017}, a meal bolus calculation method was developed for unannounced meals based on the Bergman's minimal model and unscented Kalman filter.} In Toffanin \emph{et al.} \cite{Toffanin2018}, R2R was implemented in the AP to adapt CR. 
Shi \emph{et al.} \cite{Shi2018M} explored a Bayesian optimization assisted learning {\color{v3}framework} to adapt CR profiles for the AP, using historical CGM and CSII data.

Most of the methods above improve postprandial glucose control through updating CR and CF according to the designed mechanisms. These methods do not capture the dynamics of glucose metabolism, and ignore the upcoming postprandial glucose situations in the optimization of bolus dosage, thus leading to potentially sub-optimal glycemic control. With the improved glucose sensor accuracy and accessibility, data-driven optimal control provides a promising way to achieve optimal glycemic control and reduce the burden of optimizing CR and CF. The data-driven optimal control can explicitly exploit the modeled dynamics for the bolus decision by taking account of the preprandial glucose levels, and optimizing the bolus dosage using the predicted postprandial glucose situations. This forms the motivation of our work. To implement the data-driven optimal control for the meal bolus decision, glucose prediction is a critical role. Along this direction, many methods have been explored in the literature \cite{2019Data}. {\color{v4}For example, an autoregressive with exogenous input model was presented in Romero-Ugaldein \emph{et al.} \cite{ROMEROUGALDE2019} to predict interstitial glucose. In Yu \emph{et al.} \cite{YU2018}, four different adaptive filters and a fusion mechanism were proposed for the online glucose concentration predictions.}  
Combining feature ranking with support vector regression or Gaussian processes, Georga \emph{et al.} \cite{Georga2015} investigated the short-term glucose prediction. To improve long-term glucose prediction, Montaser \emph{et al.} \cite{Montaser2020} presented an integrated predicting method based on seasonal local models and fuzzy \emph{c}-means. Different from the predictions above, using the Gaussian process (GP) regression, we provide multi-step predictions (e.g., 2 hours) for the postprandial glucose trajectories corresponding to various preprandial glucose situations, meal boluses and meal information (carbohydrate content), so that the optimization for the meal bolus can be done via the predicted glucose trajectories. 

Since GP is a data-efficient and robust modeling method, different approaches have concerned the research of GP-based control. For example, in \cite{Deisenroth2015}, a framework of probabilistic inference for learning control was proposed based on the GP and applied in real robotics and control tasks. In \cite{Achin2018}, a GP-based model predictive controller (MPC) was investigated for building energy control and demand response. {\color{v2}To enhance effective online learning and control, a risk-sensitive cost was introduced in the MPC with GP models \cite{Yang2015}, as well as in the RL \cite{Pan2018}. Inspired by the work in \cite{Yang2015, Pan2018}, we utilize the predicted information in a risk-sensitive fashion, but construct an asymmetric risk-sensitive cost with the consideration of asymmetric risks in hyper- and hypoglycemia. Finally, based on the designed cost, a constrained stochastic optimization problem is proposed for the meal bolus decision. Since the gradient of the cost cannot be computed analytically, we utilize Bayesian optimization \cite{Shahriari2015T} and Monte-Carlo method to solve the optimization problem.} For safety reasons, the final solution for the meal bolus enforces an insulin on board (IOB) constraint.    

{\color{v2}The effectiveness and robustness of the proposed method are evaluated using different \emph{in silico} protocols on the 10-adult cohort of the US Food and Drug Administration (FDA) accepted Universities of Virginia (UVA)/Padova T1DM simulator, and compared with the standard insulin bolus calculator. For the case of announced meals, the proposed method achieves satisfactory and similar performance for scenarios of nominal basal rates, in terms of mean glucose level and percent time in the safe range, without increasing the risk of hypoglycemia. Similar results are observed for the case without the meal information (assuming that the patient follows a consistent diet) and the scenarios of over/under-estimated basal rates.}
In addition, advisory-mode analysis \cite{Gillis2007} based on clinical data from a T1DM subject show that the proposed method can determine reasonable meal boluses by explicitly taking account of the preprandial glucose levels in the data-driven optimal control.
\begin{figure*}[!htb]
	\centering
	\includegraphics[width=\hsize]{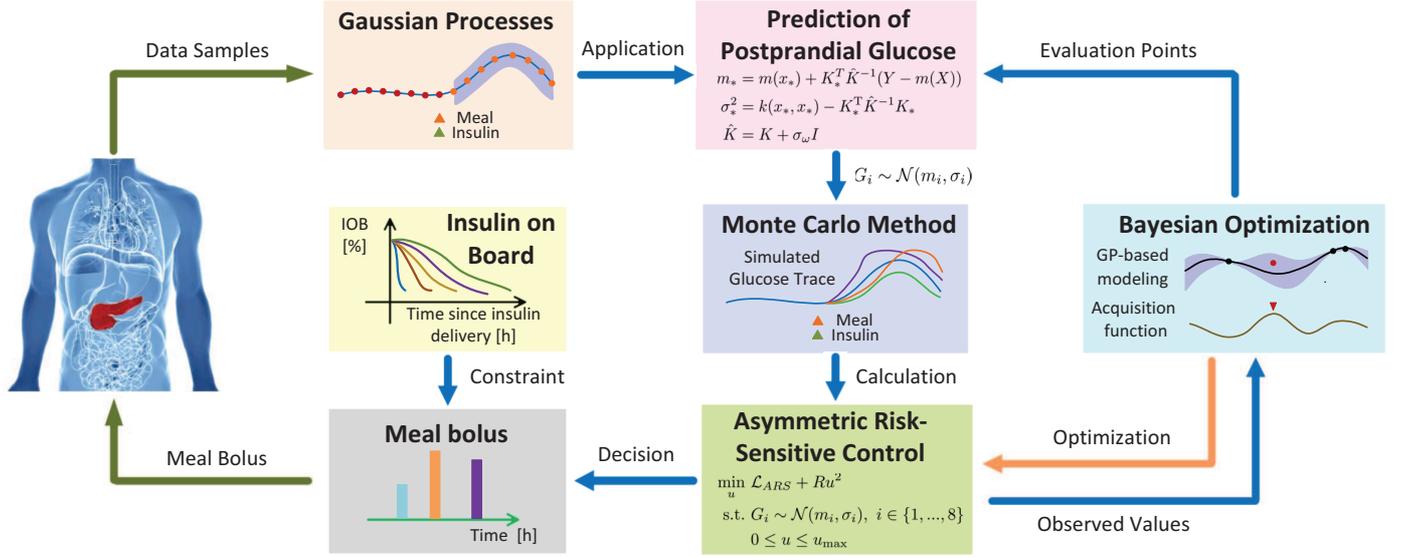}
	\caption{Schematic of the proposed method.}\label{GP-ARS}  
\end{figure*} 
\section{Materials and Methods}\label{section1}
The overall structure of the proposed data-driven meal bolus decision method is illustrated in Fig.~\ref{GP-ARS}. The method builds on three key components: model learning, asymmetric risk-sensitive control, and Bayesian optimization. 

Model learning is responsible for constructing the postprandial glucose dynamics. 
{\color{v2}Here, the aim is to provide a robust description for postprandial glucose dynamics using GPs. Specifically, fed with the  serialized data samples (including preprandial glucose measurements, the corresponding meal information, bolus dosage and postprandial glucose measurements), the GPs are trained offline and then applied  online to provide the prediction and the uncertainty estimation of postprandial glucose trajectories. Considering the asymmetric risks of hyper- and hypoglycemia and the uncertainties in the predicted glucose trajectories, we develop an asymmetric risk-sensitive cost function to favor safe  control actions. Finally, a constrained stochastic optimization problem is formulated for the meal bolus decision based on the designed cost function. Since the gradient of the cost function is unavailable, we solve the optimization problem using Bayesian optimization and Monte-Carlo simulations.} To ensure the safety of the method, IOB constraints are also incorporated.
\subsection{Gaussian Processes for Model Learning}\label{GPML}
\subsubsection{Gaussian Processes}\label{TD00}
A GP assumes a distribution over random functions $f(x): \mathbb{R}^n\rightarrow \mathbb{R}$, such that values of $f$ at any input $x$ have a joint Gaussian distribution \cite{Roberts2013, Rasmussen2006}, which is denoted as
\begin{align}
f(x)\sim \mathcal{GP}(m(x), k(x, x')),
\end{align}
where $m(x)$ and $k(x, x')$ are the mean function and positive semidefinite covariance function, respectively, which have the form: 
\begin{align}
m(x)&=\mathbb{E}_f[f(x)],\\
k(x, x')&=\text{cov}_f[f(x), f(x')],~x, x'\in\mathbb{R}^n.
\end{align}
In GPs, we can encode our prior knowledge about the process by designing corresponding prior mean and covariance functions, which are parameterized by the parameter vector $\theta_f$. Considering the noisy observations $y$ of $f$ with the form of $y=f(x)+\omega$, where $\omega\sim\mathcal{N}(0,\sigma_\omega^2)$ is a white Gaussian noise, the covariance function of $y$ has the form:
\begin{align}
k_y(x, x')=k(x,x')+\sigma_\omega^2\delta(x,x'),
\end{align}
where $\delta(x,x')$ is the Kronecker delta function which is one if and only if $x=x'$ and zero
otherwise. Given $N$ training inputs $X=[x_1,x_2,...,x_N]^\top$ and the corresponding observations $Y=[y_1,y_2,...,y_N]^\top$, the posterior GP hyper-parameters $\theta = [\theta_f, \sigma_\omega]$ are learned by maximizing the log-marginal likelihood: $\mathop{\arg\max}_{\theta} \log(p(Y|X,\theta))$ \cite{Rasmussen2006}. 

With the determined hyper-parameters, the GP can infer the posterior distribution of $y_*$ corresponding to a new input $x_*$: $y_*\sim\mathcal{N}(m_*,\sigma_*^2)$ with
\begin{align}
m_*&=m(x_*)+K_*^\top(K+\sigma_\omega I)^{-1}(Y-m(X)), \label{p1}\\
\sigma_*^2&=k(x_*,x_*)-K_*^\top(K+\sigma_\omega I)^{-1}K_*, \label{p2}
\end{align}
where $K_*=[k(x_*,x_1),...,k(x_*,x_N)]^\top$ and $K$ is the covariance matrix with elements $K_{ij}=k(x_i,x_j)$. As shown in \eqref{p1} and \eqref{p2}, for a new input, a GP can provide a determined prediction via predictive mean, c.f., \eqref{p1}, along with an estimate of uncertainty or confidence in the prediction via the predictive variance, c.f., \eqref{p2}. 

\subsubsection{Postprandial Glucose Prediction}\label{ESO}
Physiologically-based compartmental models are commonly utilized to describe glucose dynamics using first order differential equations \cite{Bergman1981, Hovorka2004, Dalla2007}; however, the parameter identification of these models is time-consuming and sometimes is even impossible with incomplete information, e.g., when only the CGM data is available. Here, we utilize GPs to perform robust modeling for postprandial glucose dynamics with incomplete and noisy information. Specifically, by feeding autoregressive, or time-delayed, input and output signals back to the model as regressors, the GPs can be used for modeling nonlinear dynamical control systems \cite{Deisenroth2015, Achin2018}.

To do this, the postprandial glucose (PG) dynamics is described in a multistep form using autoregressive models, and each step is separately represented by a nonlinear function with additive noise, which has the form:
\begin{align}
	P_{t+1} = &f_t(z_t)+w_t,\nonumber\\ z_t = &[P_{t-l},...,P_t,u,d]^\top;\nonumber\\
	P_{t+2} = &f_{t+1}(z_{t+1})+w_{t+1},\nonumber\\ z_{t+1} =& [P_{t-l+1},...,P_{t+1},u,d]^\top; \nonumber\\
	...\nonumber\\
	P_{t+n} = &f_{t+n-1}(z_{t+n-1})+w_{t+n-1}, \nonumber\\ z_{t+n-1} =& [P_{t+n-1-l},...,P_{t+n-1},u,d]^\top,
	\label{model}
\end{align}
where $t$ denotes the time of the meal intake,  $w$ is a white Gaussian noise, $P$ is the glucose measurement, and $l$ is the lag for autoregressive outputs; $u$ is the meal bolus, and $d$ is the  carbohydrate intake. To convey the most information of glucose situations, the lag of $l=7$ and the sampling period of $T= 15 \min$ are considered. This corresponds to the lag of 2 hours. {\color{v2}Correspondingly, we take $n=8$}; since the sampling period is 15 $\min$, this corresponds to the duration of 2 hours.

The utilization of GPs is divided into two stages: offline modeling and online prediction. In the offline modeling stage, based on (7), the GPs are separately used to model the PG dynamics in each step following a similar way. For example, for the time step $t+1$, we use $z_t = [P_{t-7},...,P_t,u,d]^\top$ as training inputs, and the differences $\Delta P_t = P_{t+1}-P_t$ as training targets to reduce the prediction uncertainty \cite{Deisenroth2015}. A linear mean function and a commonly-used covariance function known as squared exponential (SE) covariance kernel are considered:  
\begin{align}
m(z_t)=&a^\top z_t+b,\\
k_z(z_t, z_t')=&\sigma_f^2\exp\left(-\frac{1}{2}(z_t-z_t')^\top\Omega^{-1}(z_t-z_t')\right)+\delta(z_t, z_t')\sigma_\omega^2,\label{SE}
\end{align} 
where $\Omega=\text{diag}\{[l_1^2, l_2^2,...,l_{10}^2]\}$, $\sigma_f^2$ denotes the signal variance, and the characteristic length scales for input space $l_1, l_2,...,l_{10}$ describe the smoothness of the function. Note that $u$ and $d$ stay the same for all steps.  Moreover, to highlight the effects of $u$ and $d$ on the glucose regulation, the glucose measurements in the training inputs for each step are separately normalized into $[0,1]$ using min-max normalization. 

In the online prediction stage, given a bolus dosage and known carbohydrate amount intakes, by iteratively feeding back the predictive mean of previous step into the input for the prediction, we can obtain the prediction for each step using corresponding trained GPs. This prediction corresponds to the difference between the current step and previous step. We then add this prediction with the predicted mean of the previous step to determine the final prediction in current step. Uniting the predictions for the all steps, the GPs are able to provide 8-step predictions for the PG trajectories that correspond to the corresponding preprandial glcucose situations, carbohydrate intakes and meal boluses. 
{\color{v2}Note that when the eating habit of a subject is approximately consistent in terms of timing and sizes of meal intake,
	we can approximate the effect of similar food intake as an invariant disturbance and the meal size information can be optional for postprandial glucose prediction, utilizing the robust prediction ability of the GPs; this allows the design of a bolus decision algorithm without meal announcements.}

\subsection{Asymmetric Risk-Sensitive Control}\label{NF}
\subsubsection{Asymmetric Risk-Sensitive Cost}
{\color{v2}The glucose control problem is highly asymmetric  in the sense that consequences of hypoglycemia are immediate and more detrimental in comparison with those of (temporary) hyperglycemia; therefore, we try to correct the postprandial hyperglycemia while taking extra care of hypoglycemia. One feasible method to address this issue is to construct asymmetric cost functions \cite{Gondhalekar2016, ldg16}. Inspired by the work in \cite{ldg16}, we penalize deviations above and below the target asymmetrically; in addition, considering the uncertainties in postprandial glucose prediction, the cost function is built in the risk-sensitive (RS) framework. 
	
	To do this, we denote the 8-step predictions provided by the GPs as $G_i\sim\mathcal{N}(m_i,\sigma_i^2)$, $i\in\{1,2,...,8\}$, respectively, and collect them as a vector state $G=[G_1, G_2, ... ,G_8]^\top$, which describes the probability distribution of postprandial glucose trajectories. According to the principle of   risk-sensitive  analysis \cite{Whittle1986}, an asymmetric RS cost is   designed as follows:
	\begin{align}
	\mathcal{L}_{ARS}=-\frac{2}{\gamma}\log&\mathbb{E}\left[\exp\left(-\frac{\gamma}{2}(G-G_r)_+^\top Q^+(G-G_r)_+\right.\right.\nonumber\\
	&\left.\left.-\frac{\gamma}{2}(G-G_r)_-^\top Q^-(G-G_r)_-\right)\right],\label{ARS}
	\end{align}
	where 
	\begin{align}
	(G-G_r)_+=[(G_1-G_{r1})\mathbf{1}(G_1-G_{r1}\geq 0),...,\nonumber\\
	(G_8-G_{r8})\mathbf{1}(G_8-G_{r8}\geq 0)]^\top,\\
	(G-G_r)_-=[ (G_1-G_{r1})\mathbf{1}(G_1-G_{r1}< 0),...,\nonumber\\
	(G_8-G_{r8})\mathbf{1}(G_8-G_{r8}< 0)]^\top,
	\end{align}
	and $\mathbf{1}(\cdot)$ denotes the indicator function; $G_r$ is the target for the postprandial glucose management; $Q^+$ is a positive penalty matrix for the glucose excursions above the target; $Q^-$ is a negative penalty matrix for the glucose excursions below the target;
	$\gamma<0$ is a risk sensitivity parameter that determines the system's attitude towards uncertainty  \cite{Yang2015,Pan2018}. With the risk-sensitive cost function, the optimizer is able to fully exploit the experience learned from the historical data while keeping its own decision-making ability.

	As for the design of asymmetric penalty, $Q^+$ is designed as a constant diagonal matrix. Based on the designed $Q^+$, the diagonal elements of $Q^-$ are devised to increase exponentially with the increase of  the absolute deviation from target while being restricted by upper and lower bounds. Specifically, the $i$th diagonal element of $Q^-$ has the form of 
	\begin{align}
	Q^-_i:=Q^+_i\left(\frac{c_1}{1+\exp\{\alpha (\beta-|G_i-G_{ri}|)\}}+c_2\right),
	\end{align}
	where $Q^+_i$ is the $i$th diagonal element of $Q^+$; $\Gamma := [\alpha,\beta,c_1,c_2]$ is a quadruple determines the penalty intensity, which is designed same for the all diagonal elements.} The lower and upper bounds are determined by $c_2$ and $c_1+c_2$, respectively, and the rate of increase is controlled by $\alpha$. The parameter design will be discussed in Section~\ref{ParamDesignSec}.

Based on the above design, the quadratic penalty using constant $Q^+$ is scaled on the excursions above the target to maintain a reasonable but active response to hyperglycemia. Comparatively, a quadratic penalty with exponentially weighted coefficients is applied to the glucose excursions below the target to have a reasonably conservative response to the glucose excursions near the target, while maintaining the ability to respond quickly to larger glucose excursions and safely compensate  for hypoglycemia.
\subsubsection{GP-Based Asymmetric Risk-Sensitive Control}
Given the designed asymmetric risk-sensitive cost $\mathcal{L}_{ARS}$ in \eqref{ARS}, the GP-based asymmetric risk-sensitive control for the meal bolus decision is formulated as the following constrained stochastic optimization problem, {\color{v2} 
	\begin{align}
	\min_{u}&~\mathcal{L}(u):= \mathcal{L}_{ARS}+Ru^2,\label{NOP}\\
	\text{s.t.}& ~G_i\sim\mathcal{N}(m_i,\sigma_i),~i\in\{1,...,8\}\\
	&~0 \leq u \leq u_{\max},\label{bound}
	\end{align}}where $u$ is the meal bolus to be optimized, $R$ is the input weighting to compromise the  asymmetric RS cost and the actual needed bolus dosage, $m_i$ and $\sigma_i$ are parameters provided by the GPs. 
\subsection{Bayesian Optimization for Meal Bolus Decision}\label{BO}
{\color{v2}Since the gradient of the cost function cannot be obtained analytically, Bayesian Optimization (BO)   \cite{Shahriari2015T} is employed to solve the above constrained stochastic optimization problem.} 

\subsubsection{Model Learning for the Cost Function}\label{aei}
Here, we utilize a new GP to construct an approximation of a complex map from the decision variable $u$ to the cost function value $\mathcal{L}(u)$ in \eqref{NOP}. Specifically, we consider a prior zero mean function and the SE covariance function in \eqref{SE} with a scalar input. As discussed in Section \ref{GPML}, given the set of $N$ past observations $\mathcal{D}_{1:N}=\{u_{1:N},\mathcal{L}(u_{1:N})\}$, the GP is trained and then applied to predict the cost function value for a candidate meal bolus $u_*$ according to \eqref{p1}-\eqref{p2}. The prediction is denoted as $\hat{\mathcal{L}}(u_*)$ and will be utilized to construct the acquisition function (see Section \ref{AF}). 

Note that the value of the cost function  corresponding to the decision variable is estimated by Monte-Carlo simulations. Specifically, given a bolus dosage, we generate 1000 samples for the postprandial glucose trajectory based on the joint Gaussian distribution provided by the GP in  Section \ref{GPML}. The average cost for these samples is further calculated, which is then regarded as the observed value of the cost corresponding to the bolus dosage.

\subsubsection{Acquisition Function}\label{AF}
As a critical ingredient of the BO, the acquisition function guides the optimization by determining the optimum candidate point for the next evaluation. Specifically, utilizing the prediction information offered by the model learning phase, the acquisition function is constructed to determine the candidate point by maintain a trade-off exploration of the search space and exploitation of current promising areas. Up to now, there are rich literature concerning the acquisition function design \cite{Shahriari2015T}, where several structures have been developed, e.g., probability of improvement, expected improvement, upper confidence bound, and entropy search. In this work, the expected improvement is considered.

Compared with probability of improvement, the expected improvement (EI) acquisition function \cite{Jones1998} also incorporates the amount of improvement in selecting the candidate points. Specifically, to minimize the cost in \eqref{NOP}, the improvement function of the EI acquisition function is defined as 
\begin{align}
I(u_*):=(\mathcal{L}^p_{\min}-\hat{\mathcal{L}}(u_*))\mathbf{1}(\mathcal{L}^p_{\min}>\hat{\mathcal{L}}(u_*)),
\end{align}
where $\mathcal{L}^p_{\min}$ denotes the minimum observed value of the cost so far,  $\hat{\mathcal{L}}(u_*)\sim\mathcal{N}(m(u_*),\sigma^2(u_*))$ is the prediction at candidate point $u_*$. Note that the term $\mathcal{L}^p_{\min}-\hat{\mathcal{L}}(u_*)$ represents the amount of improvement, and the other term denotes the probability of that improvement. Based on the improvement function, the EI acquisition function is further defined as
\begin{align}
\alpha_{EI}(u_*):= \mathbb{E}(I(u_*)).\label{EI1}
\end{align}
By calculating the expectation in \eqref{EI1}, we have
\begin{align}
\label{EI2}
&\alpha_{EI}(u_*)=\nonumber\\
&\left\{\begin{array}{ll}
(\mathcal{L}^p_{\min}-m(u_*))\Phi(U)
+\sigma(u_*)\phi(U),&\textrm{if~}m(u_*)>0,\\
0,&\textrm{otherwise,}
\end{array}\right.
\end{align}
where
\begin{align}
U=\frac{\mathcal{L}^p_{\min}-m(u_*)}{\sigma(u_*)},
\end{align}
\begin{algorithm}[t]
	\caption{Bayesian optimization for meal bolus decision}
	\label{BO}
	\begin{algorithmic}[1]
		\STATE {$\mathcal{D}\leftarrow$ Initialize: $\{u_{1:8},\mathcal{L}(u_{1:8})\}$, where $u_{1:8}$ are selected equidistantly among the bound in \eqref{bound}}
		\WHILE{the number of iterations $\leq M$} 
		\STATE{Train a GP model from $\mathcal{D}$}
		\STATE{Compute the prediction at candidate $u_*$ by the GP (Eq.~\eqref{p1}-\eqref{p2})}
		\STATE{Determine candidate $u_*$ by maximizing the acquisition function (Eq.~\eqref{EI2})}
		\STATE{Observe $\mathcal{L}(u_*)$ at determined $u_*$ using the Monte-Carlo method (see Sec.~\ref{aei})} 
		\STATE{Append $\{u_{*},\mathcal{L}(u_*)\}$ to $\mathcal{D}$}
		\ENDWHILE
		\STATE {Select the input value that corresponds to the minimum observed value of the cost as the final solution}
	\end{algorithmic}
\end{algorithm}and $\Phi(\cdot)$ is the standard normal cumulative distribution function, and  $\phi(\cdot)$ denotes the standard normal probability density function. The candidate point for the next evaluation is determined as the one that maximizes $\alpha_{EI}$. 

The BO algorithm that solves the optimization problem in \eqref{NOP}-\eqref{bound} is summarized in Algorithm 1. After $M$ sequential operations of the BO, the final solution $\widetilde{u}_b$ is determined. For safety concern,  an IOB constraint \cite{Gondhalekar2018}   is enforced to prevent overbolus based on insulin delivery history. Denoting the IOB constraint as $u_{\text{IOB}}$, the final meal bolus is determined as
\begin{align}
u_b=\widetilde{u}_b-u_{\text{IOB}}.
\end{align}

\section{Data collection and Parameter design}\label{ParamDesignSec}
The parameters of the proposed method are designed and evaluated using the UVA/Padova T1DM metabolic simulator \cite{Dalla2014}. The ``average patient'' of the simulator is selected to perform the parameter design, and the obtained parameters are further evaluated using the 10 virtual adult patients. 
\subsection{Data Collection}\label{DC}
Before the parameter design and evaluation, the data samples needed for the PG model learning are collected for each patient. The samples are collected under the situation where the \emph{in silico} patients combine basal insulin and bolus insulin to control BG. To do this, a protocol with nominal basal rate and announced meals are designed for the patients. {\color{v2}The protocol starts from 5:00 on day one and lasts one week (7 days).} Considering the daily activities of diabetic patients tend to form similar patterns, e.g., with respect to meal timing and meal amount, but to mimic lifestyle disturbances, we assume the patients take breakfast, lunch, and dinner with normally distributed meal sizes (with means and standard deviations equal to [50, 75, 75] g and [3, 4, 4] g CHO) and meal times uniformly distributed in [07:00, 09:00], [11:00, 13:00], and [18:00, 20:00], respectively. The meals are all announced but the meal boluses are calculated with potentially inappropriate CR, such that the variation of the bolus is uniformly distributed in [-30\%, +30\%].
\begin{table}[!htb]
	\caption{Parameters for the proposed method}
	\centering
	\begin{tabular}{ll}  
		\hline
		Variables&Value\\\hline 
		$\gamma$&-2\\
		$R$&4\\
		$u_{\max}$&15\\
		$M$&25\\
		$Q^+$&$\text{diag}\{[0.01,\cdots,0.01,0.02,0.02]\}$\\
		$\Gamma$&$[1,10,5,1]$\\
		$G_r$&$[100,120,140,160,160,150,140,140]^\top$\\
		\hline
	\end{tabular}\label{T1}
\end{table}With the generated data by the protocol, we totally collect 7 samples for the breakfast, lunch and dinner, respectively. Each sample includes preprandial glucose measurements, corresponding bolus dosage, carbohydrate amount, and postprandial glucose measurements. {\color{v1}As the consumed CHO for lunch and dinner are similar but different from the breakfast, we construct the same PG model for both lunch and dinner, and build a separate model for the breakfast using corresponding samples, although it is also possible to build separate models for breakfast, lunch and dinner.} 

\subsection{Parameter Design}
Using the collected samples, the GPs are trained offline to provide PG predictions for the online control of the method. In the online control, the parameter design and its evaluation are performed. In the parameter design phase, a 12-hour \emph{in silico} protocol starting from 6:00 is employed, where breakfast (50 g CHO), lunch (75 g CHO) are consumed at 8:00 and 12:00, respectively.

The parameters are designed using a trial-and-error approach based on the glucose data obtained from the ``average patient'', such that satisfactory postprandial regulation performance in terms of average glucose, percent time in [70, 180] mg/dL, and percent time below 70 mg/dL can be achieved. The obtained parameters are summarized in Table \ref{T1}. In the evaluation phase, different scenarios are designed to evaluate the performance of the proposed method with the obtained parameters. The results are reported in Section~\ref{spa}.


\section{Performance Analysis}\label{spa}
As mentioned above, the proposed method is evaluated on the 10-adult cohort of the UVA/Padova T1DM simulator. Besides, the advisory-mode analysis \cite{Gillis2007} is performed for the method based on the clinical data from a particular T1DM subject who undertook the MDI therapy. 

For the \emph{in silico} evaluations, two protocols are designed to perform the comparison of the proposed method and the standard insulin bolus calculator (denoted as ``Control''), {\color{v1}which has the form:
	\begin{align}
	u_{\text{bolus}} = \frac{\text{CHO}}{\text{CR}}+\frac{G_{c}-G_{sp}}{\text{CF}}-u_{\text{IOB}},
	\end{align}
	where $G_{c}$ is the current glucose level (mg/dL); {\color{v2}$G_{sp}$ is the glucose set-point (mg/dL), selected as $140$ in this comparison. One of the designed protocol (denoted as ``Protocol A'') begins at 5:00 on day 1 and lasts two days (48 hours) where breakfast (55 g and 45 g CHO), lunch (65 g and 85 g CHO) and dinner (85 g and 65 g CHO) are consumed at 8:00, 12:00 and 18:00 for two days, respectively. The other protocol (denoted as ``Protocol B'') begins at 5:00 and lasts one day (24 hours), including breakfast, lunch and dinner with 45 g, 85 g, and 65 g of CHO content at 8:00, 12:00 and 18:00, respectively.} Based on protocol A, \emph{in silico} experiments are performed on the whole 10-adult cohort with nominal basal rates using additive CGM noises for scenario (not) utilizing the meal information for the prediction, thus a total of 10 simulations are performed for each scenario and each method. The statistical results are summarized in Table~\ref{T2}, and the comparison performance in terms of glucose regulation and meal bolus is illustrated in Fig.~\ref{ameal}, where the 5\%, 25\%, 75\% and 95\% quartile curves together with the median curves are presented. Besides, to illustrate the robust decision-making ability of the proposed method, \emph{in silico} evaluations without meal information for scenarios of basal-rate mismatches (110\% and 80\% of the nominal basal rate) are performed on the same cohort using protocol B, respectively. The results are presented in Table~\ref{T3} and Fig.~\ref{uob}, and the corresponding discussions for the both protocol are provided in Section~\ref{A}.
	\begin{table*}[!htb]
		\centering
		\caption{Comparison results with nominal basal rate}
		\label{T2}
		\begin{tabular}{lcclccl}
			\hline
			\hline
			Scenario (\#Simulations=10)&\multicolumn{3}{c}{With meal information}&\multicolumn{3}{c}{Without meal information}\\\cmidrule(lr){2-4}\cmidrule(lr){5-7} Metric &Control &Proposed &$p$ value &Control  &Proposed &$p$ value \\ \hline
			\% time&\\
			~~~~~~$<$54 mg/dL& 0.0 (0.0) &0.0 (0.0) &1.000& 0.0 (0.0) &0.0 (0.0) &1.000\\
			~~~~~~$<$70 mg/dL& 0.0 (0.0) &0.0 (0.0) &0.500& 0.0 (0.0) &0.0 (0.0) &0.500\\
			~~~~~~70-180 mg/dL& 95.8 (5.7) &96.6 (7.1) &0.627 & 95.8 (5.7) &95.4 (7.8) &0.264\\
			~~~~~~$>$180 mg/dL& 4.2 (5.7) &3.4 (5.6) &0.605 & 4.2 (5.7) &3.7 (7.8) &0.223\\
			~~~~~~$>$250 mg/dL& 0.0 (0.0) &0.0 (0.0) &1.000& 0.0 (0.0) &0.0 (0.0) &1.000\\
			Mean glucose (mg/dL)& 132.8 (8.3) &\textbf{126.7} \textbf{(13.3)} &\textbf{0.037}&132.8 (8.3) &131.0 (10.9) &0.084\\
			SD glucose (mg/dL)& 26.2 (7.2) &25.9 (8.9) &1.000& 26.2 (7.2) &\textbf{27.0} \textbf{(5.8)} &\textbf{0.020}\\
			Mean glucose at 07:00& 118.8 (11.0) &\textbf{116.5} \textbf{(9.5)} &\textbf{0.008} &118.8 (11.0) &117.3 (7.0) &0.172\\
			\hline
			\multicolumn{7}{l}{Data in this table are shown as median (inter quartile range), and $p$ values are calculated based on the number}\\
			\multicolumn{7}{l}{of virtual patients. Statistically significant ($p < 0.05$) changes are highlighted in bold.}
		\end{tabular}
	\end{table*}
	\begin{figure*}[!htb]
		\centering
		\subfigure[With meal information]
		{\includegraphics[width=0.49\textwidth]{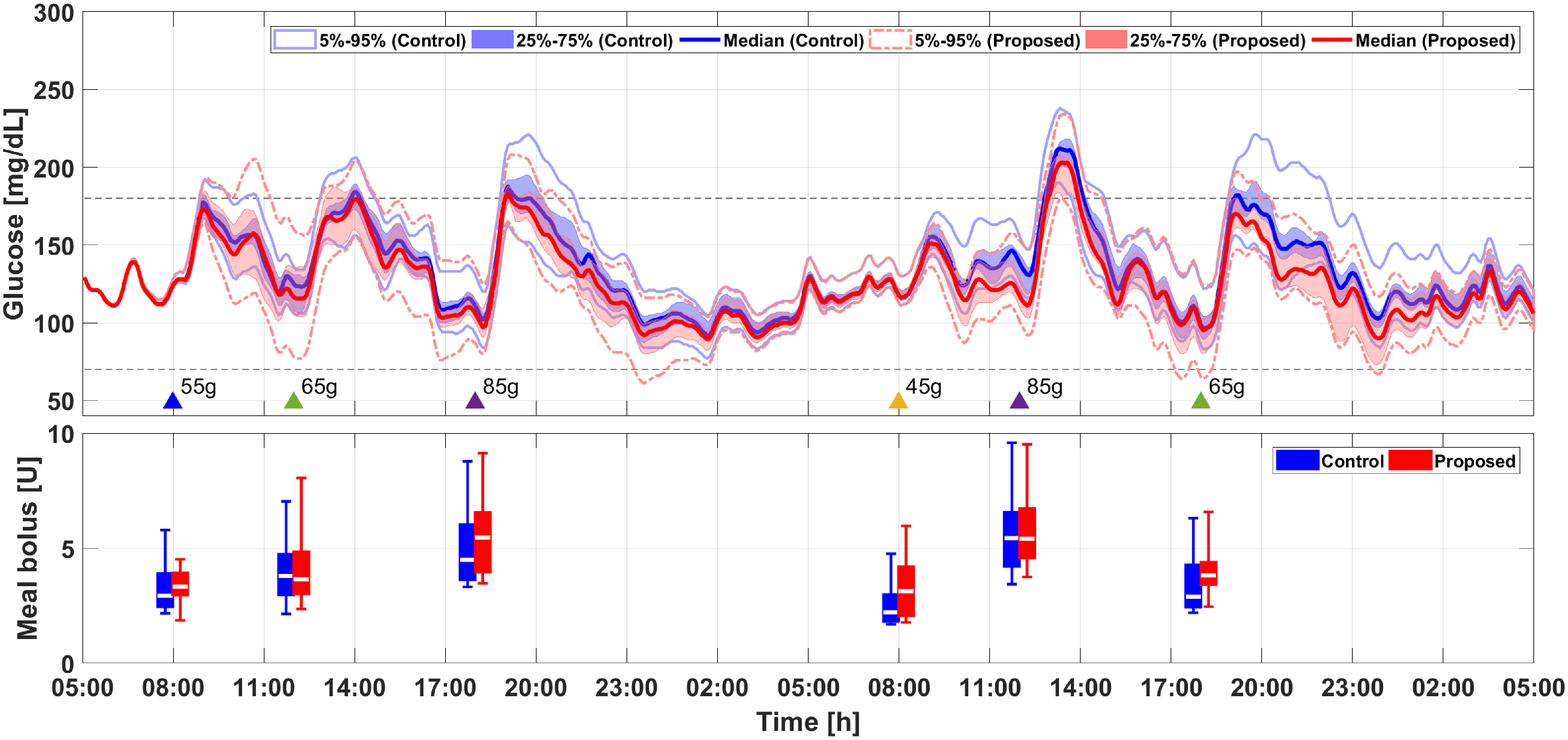}}
		\subfigure[Without meal information]
		{\includegraphics[width=0.49\textwidth]{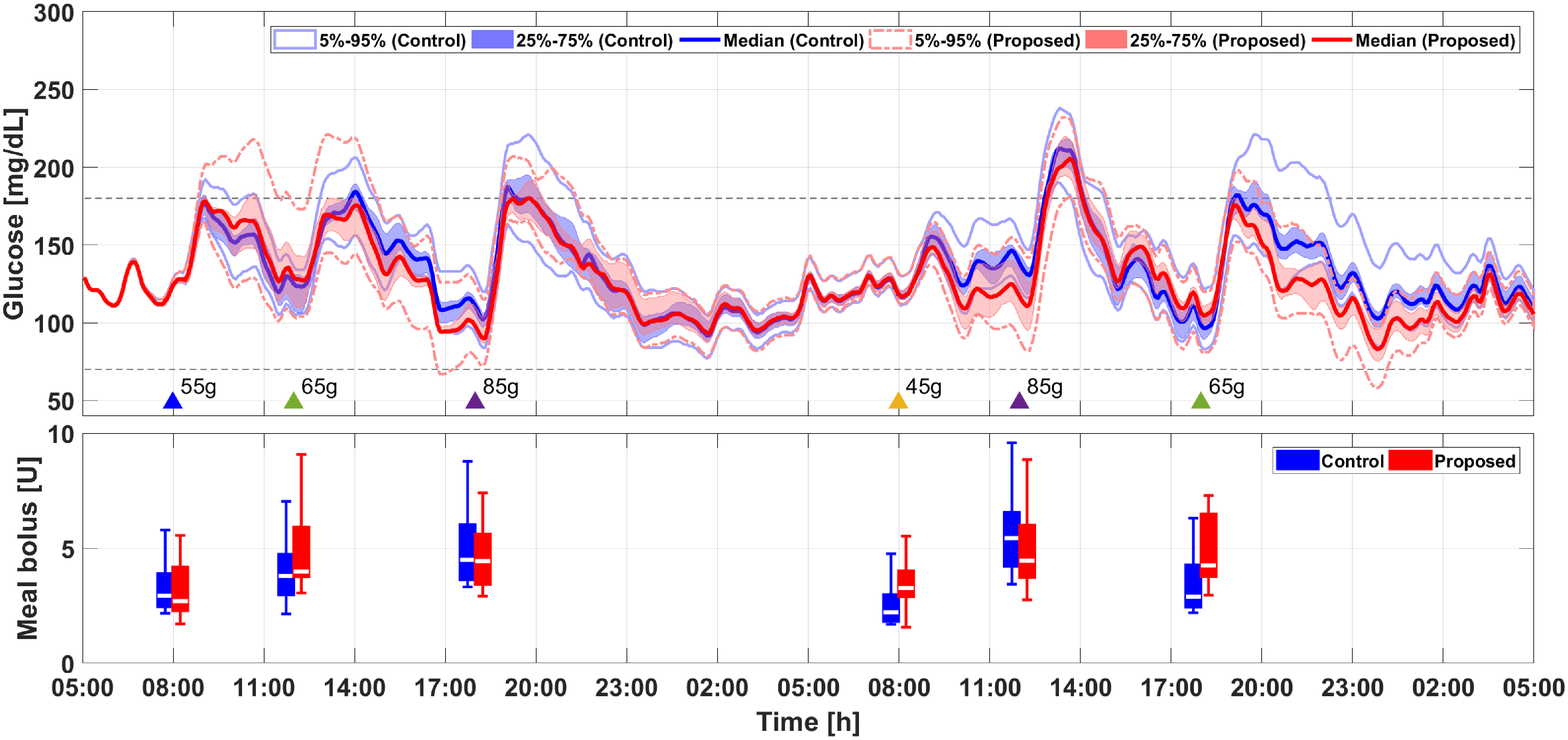}}
		\caption{Performance comparison with nominal basal rate in terms of glucose regulation and meal bolus. Yellow, blue, green and purple triangles denote meals of 45 g, 55 g, 65 g and 85 g CHO, respectively.}\label{ameal}
	\end{figure*}
		\begin{table*}[!htb]
	\centering
	\caption{Comparison results with under/over-estimated basal rate}
	\begin{tabular}{lcclccl}
		\hline
		\hline
		Scenario (\#Simulations=10)&\multicolumn{3}{c}{80\% of the nominal basal rate}&\multicolumn{3}{c}{110\% of the nominal basal rate}\\\cmidrule(lr){2-4}\cmidrule(lr){5-7} Metric &Control &Proposed &$p$ value &Control  &Proposed &$p$ value \\ \hline
		\% time&\\
		~~~~~~$<$54 mg/dL& 0.0 (0.0) &0.0 (0.0) &1.000& 0.0 (0.0) &0.0 (0.0) &1.000\\
		~~~~~~$<$70 mg/dL& 0.0 (0.0) &0.0 (0.0) &1.000& 0.0 (0.0) &0.0 (3.1) &0.625\\
		~~~~~~70-180 mg/dL& 87.8 (3.5) &87.7 (8.0) &0.389& 96.5 (3.8) &95.0 (9.7) &0.586\\
		~~~~~~$>$180 mg/dL& 12.2 (3.5) &12.3 (8.0) &0.389& 3.5 (4.9) &4.7 (7.3) &0.945\\
		~~~~~~$>$250 mg/dL& 0.0 (0.0) &0.0 (0.0) &1.000& 0.0 (0.0) &0.0 (0.0) &1.000\\
		Mean glucose (mg/dL)& 151.2 (4.9) &146.6 (10.4) &0.131& 120.9 (6.9) &119.0 (7.6) &0.064\\
		SD glucose (mg/dL)& 24.7 (5.9) &26.8 (5.9) &0.131& 28.8 (4.8) &32.0 (8.9) &0.275\\
		Mean glucose at 07:00& 151.0 (21.0) &\textbf{145.0} \textbf{(27.0)} &\textbf{0.047}& 93.0 (11.0) &93.0 (12.0) &0.559\\
		\hline
		\multicolumn{7}{l}{Data in this table are shown as median (inter quartile range), and $p$ values are calculated based on the number}\\
		\multicolumn{7}{l}{of virtual patients. Statistically significant ($p < 0.05$) changes are highlighted in bold.}
	\end{tabular}\label{T3}
\end{table*}
\begin{figure*}[!htb]
	\centering
	\centering
	\subfigure[Under-estimated basal rates]
	{\includegraphics[width=0.49\textwidth]{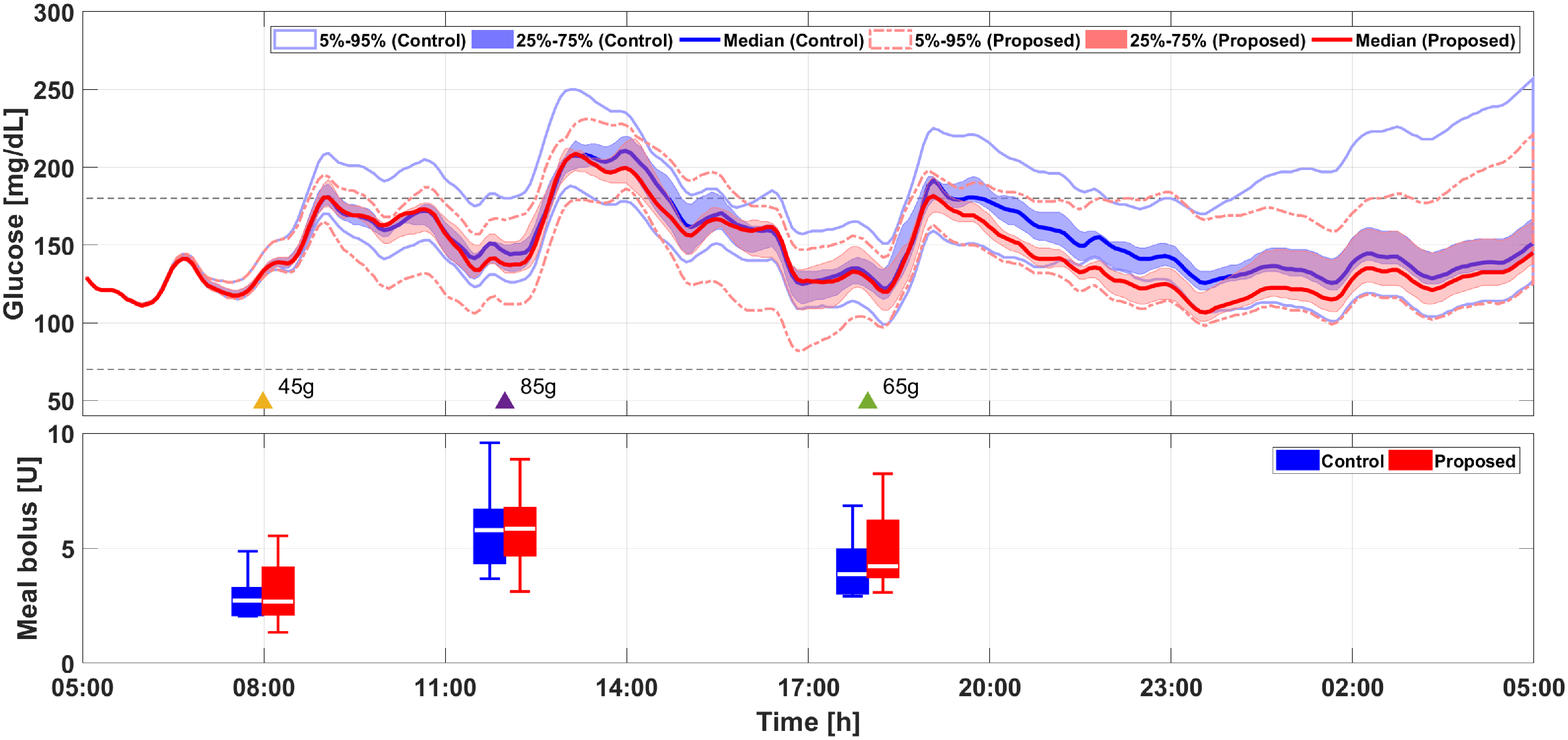}}
	\subfigure[Over-estimated basal rates]
	{\includegraphics[width=0.49\textwidth]{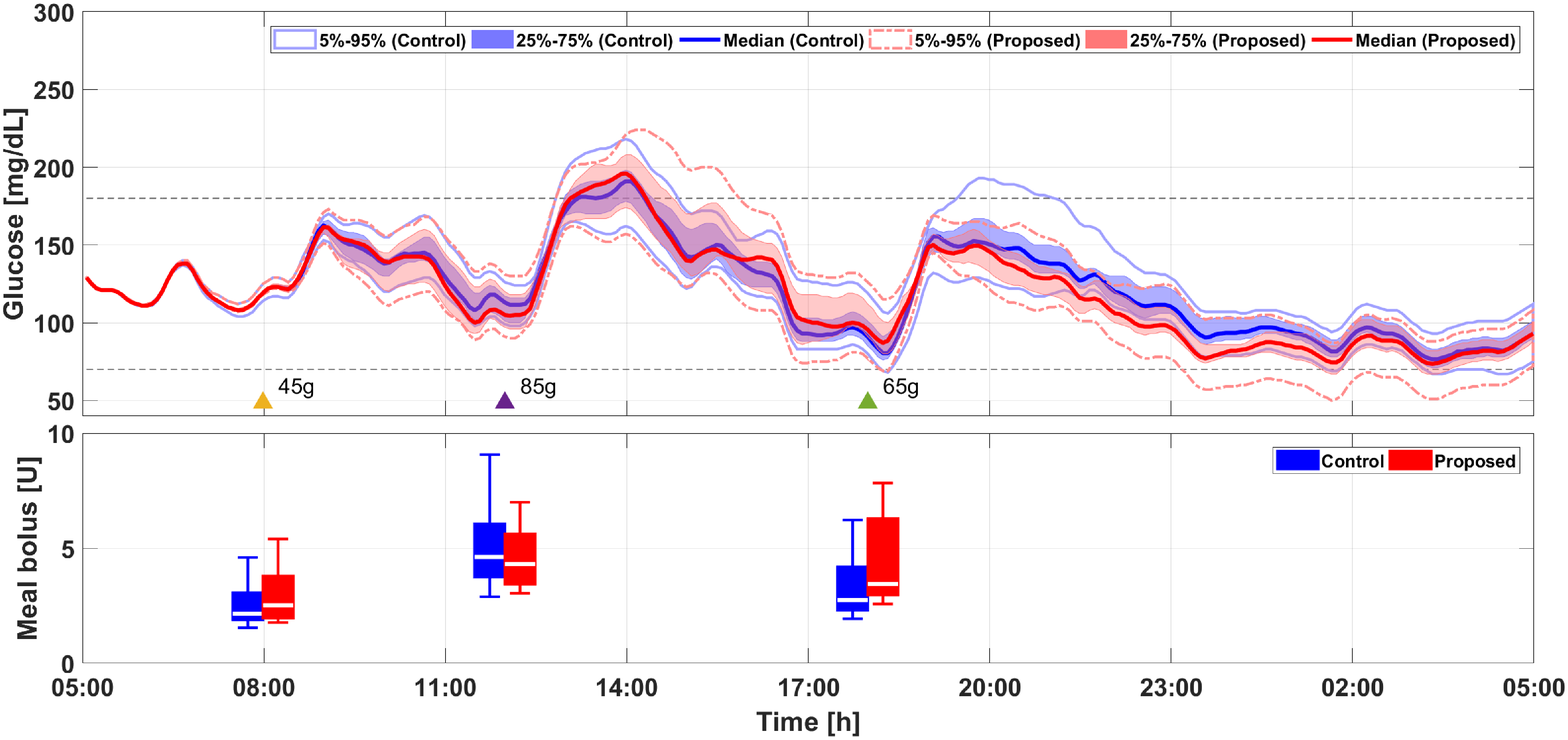}}
	\caption{Performance comparison for the scenario of under-/over-estimated basal rates.}
	\label{uob}
\end{figure*}
	
	{\color{v1}The advisory-mode analysis \cite{Gillis2007} allows the comparisons with insulin recommendations made by clinicians, through feeding the identical glucose data obtained in the clinical trial to the system. Here, by feeding the historical preprandial glucose data to the proposed method, the corresponding meal boluses are determined and compared with the ones following the clinician's advice. The obtained meal boluses have no causal impact on historical data and are only for the comparison. The collection of the historical clinical data are presented in Fig.~\ref{clinical_data}, and the comparison performance are illustrated in Fig.~\ref{clinical_evaluation}. The corresponding discussions are provided in Section~\ref{C}.}
	
	\subsection{In Silico Performance Evaluation}\label{A}
	In this subsection, using Protocol A, the performance of the proposed method are evaluated for the cases with the meal information (IM) and without the meal information (NM), respectively. From Table~\ref{T2}, for the both cases, {\color{v2}the proposed method achieves comparable glucose regulation performance in comparison with the standard insulin bolus calculator, which is equipped with the well-designed CR and CF. This is reflected in percent time in the euglycemic range of 70-180 mg/dL (96.6\% vs. 95.8\%, $p=0.627$ for IM; 95.4\% vs. 95.8\%, $p=0.264$ for NM), mean glucose (126.7 mg/dl vs. 132.8 mg/dL, $p=0.037$ for IM; 131.0 mg/dL vs. 132.8 mg/dL, $p=0.084$ for NM), percent time $>$ 250 mg/dL (0.0\% vs. 0.0\%,  $p=1.000$ for IM; 0.0\% vs. 0.0\%,  $p=1.000$ for NM) and glucose standard deviation (SD) (25.9 vs. 26.2, $p=1.000$ for IM; 27.0 vs. 26.2,  $p=0.020$ for NM). No increase in the risk of hypoglycemia is observed (percent time $<$ 70 mg/dL, 0.0\% vs. 0.0\%,  $p=0.500$ for IM; 0.0\% vs. 0.0\%,  $p=0.500$ for NM). These results show the method is robust to the meal information if the CHO amount intakes are similar to the standard amount used for the model learning}, and {\color{v1}also illustrate the learning ability of proposed method, as the samples for the model learning are collected with the inappropriate CR.} The discussions of control performance are consistent with the quartile curves in Fig.~\ref{ameal}. Besides, from the quartile curves of meal bolus in Fig.~\ref{ameal}, we observe that compared with the case of NM, the method in the case of IM tends to increase (decrease) the meal bolus for the known large (small) CHO. This implies that when the accurate meal information are available, the method can react reasonably to the CHO amount, but this will increase the burden of data collection for model learning in return.
	
	{\color{v2}Using Protocol B, we also perform additional tests considering realistic scenarios of under/over-estimated basal rates with/without meal information to evaluate the robustness of the proposed method. The results are also compared with those obtained for the standard insulin bolus calculator. Since  similar results are observed for   both cases, here we only present the results of the case without meal information.} From Table~\ref{T3} and Fig.~\ref{uob}, it is observed that for the scenario of under-estimated basal rate, the proposed method tends to increase the meal bolus for the elevated preprandial glucose levels, and {\color{v2}achieves similar performance in terms of percent time in [70, 180] mg/dL (87.7\% vs. 87.8\%, $p$ = 0.389) and mean glucose (146.6 mg/dL vs. 151.2 mg/dL, $p=0.131$) without causing risk of hypoglycemia (percent time below 70 mg/dL, 0.0\% vs. 0.0\%,  $p=1.000$). Similar results are observed for the scenario of over-estimated basal rate, which are reflected in percent time in [70, 180] mg/dL (95.0\% vs. 96.5\%, $p$ = 0.586) and mean glucose (119.0 mg/dL vs. 120.9 mg/dL, $p=0.064$) without causing risk of hypoglycemia (0.0\% vs. 0.0\%,  $p=0.625$).} 
	These results further illustrate the robust decision-making ability of the method for the extreme preprandial glucose situations.
	
	Finally, we would like to note that the extremely satisfactory  glucose control performance obtained by the proposed approach and the standard bolus calculator is partially attributed to the simulator dynamics and should not be over-emphasized, as we observe that the time-in-range achieved using the standard bolus calculator (using the CR and CF values provided by the simulator) goes beyond $95\%$ in Table~\ref{T2}.   The aim of presenting the \emph{in silico} evaluation results, however, is to compare the proposed data-driven method with the standard approach that is built on CR and CF information. 
\begin{figure*}[!htb]
	\centering
	\includegraphics[width=0.8\hsize]{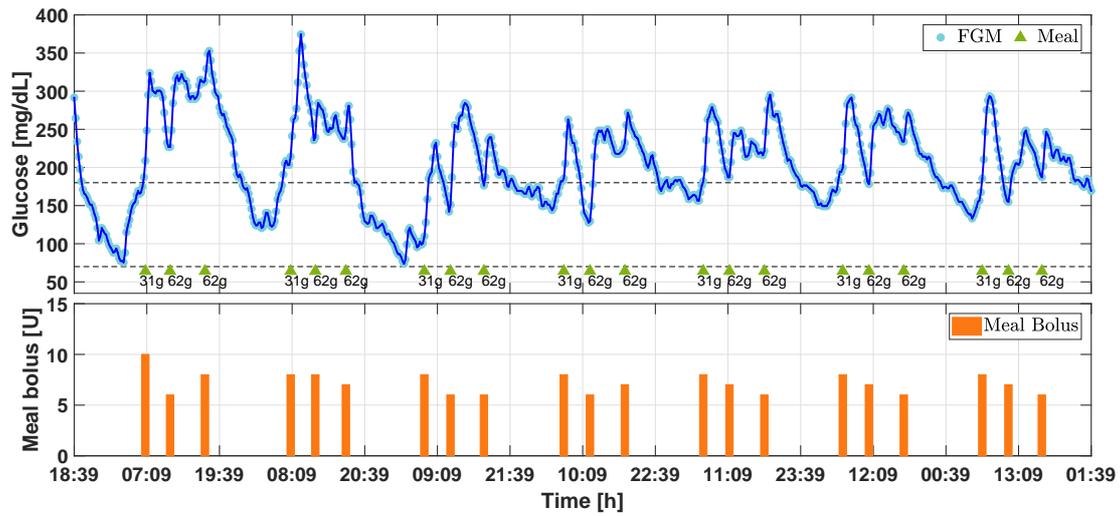}
	\caption{The collection of the clinical data from a T1DM subject. Meals are denoted by green triangles with sizes below them, and the corresponding meal boluses determined by the clinicians are displayed in the second panel.}\label{clinical_data}  
\end{figure*}
\begin{figure*}[!htb]
	\centering
	\includegraphics[width=0.8\hsize]{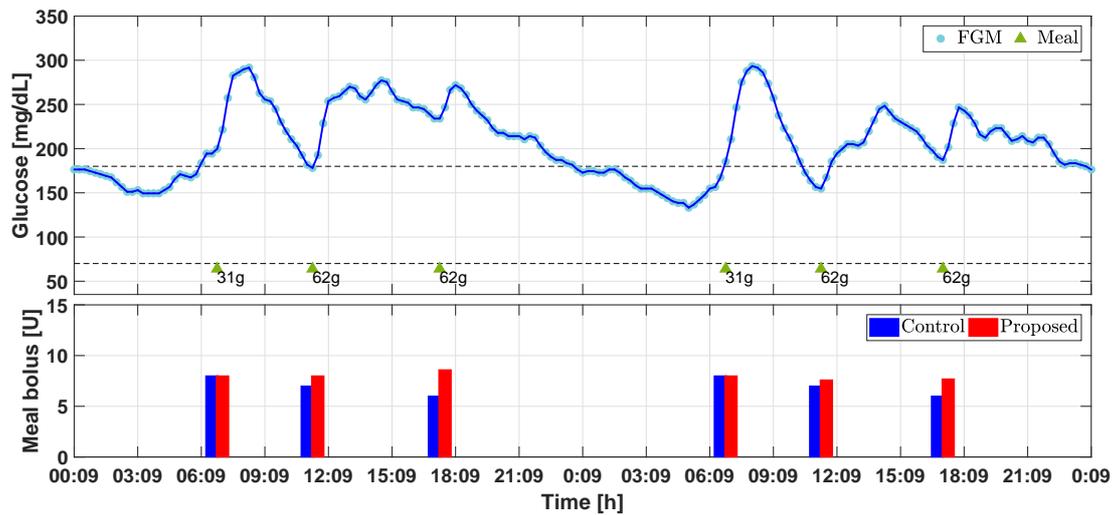}
	\caption{Performance evaluation of the proposed method based on the clinical data. Meals are
		denoted by green triangles with sizes below them. }\label{clinical_evaluation}  
\end{figure*}
	\subsection{Advisory-Mode Comparisons Using Clinical Data}\label{C}
	{\color{v1}In this subsection, the historical clinical data from a T1DM subject who undertook the MDI therapy are utilized to evaluate the performance of the proposed method.} {\color{v2}Flash glucose monitoring (FGM) was worn to collect the glucose measurements. The data for the seven days (see Fig.~\ref{clinical_data}) were collected at hospital, where the patient was managed to have a consistent diet for every day, in terms of similar meal timing and meal intakes, and the corresponding meal boluses were determined by the clinicians.} {\color{v1}The study was approved by institutional review board at Peking University People's Hospital and written informed consent of the participant was obtained.} {\color{v2}Since the meal intakes are almost identical for every day, we use the data of the first five days to model the PG dynamics for the breakfast and lunch-dinner, respectively, without the meal information.} At last, the performance of the proposed method equipped with the trained GPs is evaluated by feeding the data for the next two days. 
	
	From Fig.~\ref{clinical_evaluation}, compared with the fixed meal boluses (denoted as blue bars) following the clinician's advice, the proposed method can determine reasonable meal boluses (denoted as red bars) according to the preprandial glucose levels. {\color{v2}We observe that the method would suggest additional 1-2 units of the insulin bolus for the lunch or dinner of the two days due to the elevated preprandial glucose level. This reasonable increase bolus would help reduce the later happened hyperglycemia. Besides, the breakfast meal boluses  are the same as those determined by the clinicians; observing that the historical breakfast boluses are almost all identical, this indicates that the risk-sensitive control mechanism tends to maintain the decisions in the database to ensure safety when the risk of taking a different value is not clear.} 
	
	\section{Conclusion}
	In this work, a GP-based asymmetric risk-sensitive (ARS) control method is proposed for the personalized meal bolus decision. With the formulation of the ARS cost function, the method is capable to apply the experience learned form the samples, while keeping own decision-making ability, e.g., taking extra care of the hyperglycemia and increasing (decreasing) the meal bolus for the elevated (lowered) preprandial glucose levels. {\color{v2}Besides, the method is robust to the meal variability within a tolerable range, which reduces the burdens of estimating the CHO amount for each meal.} The effectiveness and robustness of the controller are evaluated using the 10-adult cohort of the UVA/Padova simulator through comparisons with the standard insulin bolus calculator. Also, advisory-mode analysis is performed based on the clinical data from a T1DM subject. For future work, since the model learning is relied on the offline samples, an adaptive PG model will be developed for the method to take the long-term variations of the physiological dynamics into account. 

	\section*{Acknowledgment}
	Access to the distributed version of the University of
	Virginia (UVA)/Padova metabolic simulator for research
	purposes is acknowledged.

	%

	%
	%
	
 \bibliographystyle{elsarticle-num} 
 \bibliography{Caidh}




\end{document}